\documentclass[manuscript]{aastex}

\shorttitle{Heavy Element Enrichment of a Jupiter-mass Protoplanet}
\shortauthors{Helled \& Schubert}

\begin{document}


\title{Heavy Element Enrichment of a Jupiter-mass Protoplanet as a Function of Orbital Location}


\author{R. Helled\altaffilmark{1} and G. Schubert \altaffilmark{1}}
\affil{Department of Earth and Space Sciences and Institute of Geophysics and Planetary Physics,\\
University of California, Los Angeles, CA 90095 1567, USA}

\email{rhelled@ess.ucla.edu (R. Helled); schubert@ucla.edu (G. Schubert)}

\begin{abstract}
One possible mechanism for giant planet formation is disk instability in which the planet is formed as a result of gravitational instability in the protoplanetary disk surrounding the young star. 
The final composition and core mass of the planet will depend on the planet's mass, environment and the planetesimal accretion efficiency. We calculate heavy element enrichment in a Jupiter-mass protoplanet formed by  disk instability at various radial distances from the star, considering different disk masses and surface density distributions. Although the available mass for accretion increases with radial distance (a) for disk solid surface density ($\sigma$) functions $\sigma=\sigma_0 a^{-\alpha}$ with $\alpha <$ 2, the accretion timescale is significantly longer at larger radial distances. Efficient accretion is limited to the first $\sim 10^5$ years of planetary evolution, when the planet is extended and before gap opening and type II migration take place. The accreted mass is calculated for disk masses of 0.01, 0.05 and 0.1 M$_{\odot}$ with $\alpha =$ 1/2, 1, and 3/2.
We show that a Jupiter-mass protoplanet can accrete 1 to 110 M$_{\oplus}$ of heavy elements, depending on the disk properties. Due to the limitation on the accretion timescale, our results provide lower bounds on heavy element enrichment. Our results can explain the large variation in heavy element enrichment found in extra-solar giant  planets.  Since higher disk surface density is found to lead to larger heavy element enrichment, our model results are consistent with the correlation between heavy element enrichment and stellar metallicity.  
Our calculations also suggest that Jupiter could have formed at a larger radial distance than its current location while still accreting the mass of heavy elements predicted by interior models.
We conclude that in the disk instability model the final composition of a giant planet is strongly determined by its formation environment. The heavy element abundance of a giant planet does not discriminate between its origin by either disk instability or core accretion. 
\end{abstract}



\section{Introduction}
Jupiter contains mostly hydrogen and helium with a smaller fraction of heavy elements. Constraints on the mass of Jupiter's heavy elements can be found by fitting its measured gravitational field using theoretical equations of state. Theoretical models of Jupiter's interior predict that the total mass of heavy elements ranges from 20 to 40 M$_{\oplus}$ (Saumon and Guillot, 2004; Nettelmann et al., 2008; Militzer et al., 2008). In addition, measurements by the Galileo probe mass spectrometer suggest that Jupiter's atmosphere is enriched in heavy elements by a factor of up to three compared to the Sun (Young, 2003). 
Observations of transiting giant planets suggest that the composition of gas giant planets varies over a wide range. While some extra-solar planets are found to contain mainly hydrogen and helium others possess large amounts of heavy elements (Guillot, 2007). \par

Giant planet formation models must be able to explain the abundances of heavy elements in Jupiter and other giant planets. The standard model for giant planet formation is 'core accretion' (Pollack et al., 1996; Hubickyj et al., 2005; Lissauer, \& Stevenson, 2007) in which formation starts with the build-up of a solid core. Once the core reaches a critical mass of roughly 10 M$_{\oplus}$, runaway gas accretion occurs leading to the accumulation of a massive gaseous envelope. Until recently, the formation timescale in this model was uncomfortably long compared to observationally derived lifetimes of protoplanetary disks (Haisch et al., 2001). However, recent work offers several ways to overcome the 'timescale problem' in the core accretion model (e.g., Hubickyj et al., 2005; Doson-Robinson et al., 2008; Lissauer et al., 2009); one of them is planetary migration (Alibert et al., 2005; Dodson-Robinson et al., 2008). Observations of 'hot Jupiters' have supported the paradigm in which planets are formed at large radial distances and then migrate inward to their present locations. Alibert et al. (2005) have shown that Jupiter can form within $\sim$ 1 Myr starting with a planetary embryo at 8 AU. \par

An alternative model for giant planet formation is disk instability (Boss, 1998; Mayer et al., 2002). In this model giant planets are formed as a result of gravitational instabilities in the protoplanetary disk surrounding the young star. If the  formed 'clumps' cool fast enough to stay gravitationally bound (Rafikov, 2007; Boss, 2009) they can evolve to become gaseous protoplanets (Durisen et al., 2007).  
Gravitational instabilities can occur at very large radial distances of tens of astronomical units. In fact, formation of a clump farther from the protostar in this model is favored due to lower temperatures and faster cooling in the outer regions. Therefore, if giant planets are formed via disk instability, it is certainly possible that they are formed farther out and migrate inward to their present locations.\par 

In the disk instability model protoplanets are initially formed with a stellar composition; they can be enriched in heavy elements after their formation as a result of planetesimal accretion. Helled et al. (2006), hereafter paper I,  presented a calculation of heavy element enrichment in a Jupiter-mass clump, taking into account different planetesimal sizes, compositions, and velocities. The location of the protoplanet, however, was limited to 5.2 AU, Jupiter's current location. 
Observations of close-orbiting extra-solar giant planets suggest that planetary migration may be a common phenomenon. In addition, the recent direct imaging of extra-solar planets has revealed that giant planets can be present at much larger radial distances than that of Jupiter in the solar system (Marois et al., 2008). \par 
In this paper, we investigate possible heavy element enrichment of a Jupiter-mass clump formed at different radial distances and disk environments (masses and density profiles). The implications of the results for Jupiter and extra-solar giant planets are discussed. 

\section{Planetesimal Accretion}
The amount of solids available for accretion by a protoplanet depends on the disk mass and its radial density profile. 
The total (gas and dust) surface density $\Sigma$ profile can be given by
\begin{equation}
\Sigma(a)=\Sigma_0\left(\frac{a}{5 AU}\right)^{-\alpha},
\end{equation}   
where $\Sigma_0$, the surface density at 5 AU, is determined by the disk mass, $a$ is the radial distance, and $\alpha$ is a parameter which defines the steepness of the density function. We take the disk's inner and outer boundaries to be 0.1 and 30 AU, respectively. The value for the inner boundary is typically taken as 0.1 AU (Andrews and Williams, 2007) since disks are expected to be truncated by the stellar magnetosphere in the innermost regions (Shu et al., 1994). Indeed, observations suggest truncation radii of about 0.1 AU (Eisner et al., 2005). The value of the inner boundary radius, however, is unimportant as long as it is significantly smaller than the outer boundary radius (Ruden and Lin, 1986). A disk which ranges up to 30 AU, as considered here, covers the region in which giant planets are likely to form, and allows surface densities that are consistent with planet formation. Disks are expected to expand with time, and observations of disks suggest that they can be as large as hundreds of AU (Andrews and Williams, 2007); however, our calculation concentrates on the first $10^5$ years, before the disk expands to such large distances (Dodson-Robinson et al., 2008). \par

The relation between the disk mass and the surface density at 5 AU $\Sigma_0$ is then given by,
\begin{equation}
\Sigma_0 = \frac{M_{disk}(2-\alpha)}{2 \pi 5^{\alpha} AU^2} \left(30^{2-\alpha} -0.1^{2-\alpha} \right)^{-1},
\end{equation}  
where the disk mass $M_{disk}$ is given by $M_{disk} = \int_{0.1 AU}^{30 AU} 2\pi a \Sigma(a) da$. 
The solid surface density $\sigma$ is taken to be $\Sigma$/70
, as commonly taken for the solar nebula (Weidenschilling, 1977). Since the masses and density distributions of protoplanetary disks have wide ranges (Eisner et al., 2008; Eisner and Carpenter, 2006;  Andrews and Williams, 2007; Andrews et al., 2008), we consider three different disk masses, 0.1, 0.05, and 0.01 M$_{\odot}$, with M$_{\odot}$ being the Sun's mass, and take $\alpha$ values of 1/2, 1, and 3/2. Figure 1 shows the solid surface density for the cases considered. In all cases the surface density decreases with radial distance, however, the density decreases more slowly with radius for smaller $\alpha$ values. Table 1 summarizes the solid surface density at 5 AU ($\sigma_0$) values for the different disk masses and density profiles. Since we take the solid surface density as a constant fraction of the gas density (constant metallicity), it increases with increasing disk mass. However, high solid surface densities can also be the result of high metallicity. It is therefore possible to have low mass disks with relatively high solid surface densities, or low solid surface densities for massive disks, for high-metallicity and low-metallicity disks, respectively. The implications of high solid surface density as a result of stellar metallicity are discussed in section 3.2. \par

The mass of solids accreted by a protoplanet depends on the solid surface density (the available mass at a certain location) and  the protoplanet's efficiency in capturing solid planetesimals. 
To compute the planetesimal capture rate, we model the evolution of a Jupiter-mass protoplanet, with an initial state as expected in the disk instability model, similar to the one used in paper I. The evolution is followed assuming the protoplanet is spherical, hydrostatic and isolated (see paper I and Helled et al. (2008) for further details). The effect of planetesimal accretion is not included in the evolution calculation. Figure 2 presents the physical parameters of the protoplanet as a function of time for the first $10^5$ years. \par

As in paper I, we follow planetesimal trajectories with increasing impact parameters in order to determine the largest impact parameter for which planetesimal capture is possible $b_{capture}$. 
The computed trajectory accounts for gas drag and gravitational forces, assuming that the gravitational interaction is two-body (Podolak et al., 1987, paper I). 
The protoplanet's capture radius $R_{capture}$ is defined as the planetesimal's closest distance of approach to the protoplanet's center for the critical impact parameter (see paper I for details). Both the protoplanet radius and capture radius decrease as the protoplanet evolves. The difference between the protoplanet radius and capture radius as a function of time can be seen in figure 2. The capture radius is smaller than the actual radius of the planet especially at the early stages of the evolution, when the density in the outer envelope is low, and the planetesimal must pass through more dense inner regions to lose kinetic energy by gas drag and be captured by the protoplanet. The capture radius is typically smaller by a factor of a few than the largest impact parameter for capture (paper I). \par

Once the capture radius is known the capture cross-section can be computed. 
A 2-body gravitational cross section is given by $\pi b_{capture}^2$, however, this cross-section accounts only for the planetesimal and the protoplanet. To account for the influence of the Sun, we follow paper I and take the capture cross-section to be $\pi R_{capture}^2F_g$, as expected for a 3-body interaction (Greenzweig and Lissauer, 1990).
The gravitational enhancement parameter $F_g$ is taken as unity, providing a conservative estimate for the capture cross-section.
The planetesimals are taken to be 1 km radius bodies composed of a mixture of ice, rock, and organics with an average density of 2 g cm$^{-3}$,  and a random velocity of 1 km s$^{-1}$. Once the capture cross-section is known, the accretion rate can be computed. \par

The planetesimal accretion rate is given by (Safronov, 1969)
\begin{equation}
\frac{dm(a)}{dt}=\pi R^2_{capture}(t) \sigma(a,t) \Omega(a)
\end{equation}
where $\sigma $ is the surface density of solid
material, and $\Omega $ is the protoplanet's orbital frequency.\par

The total available mass of solids in the planet's feeding zone is given by
\begin{equation}
M_{av}(a)=\pi (a_{out}^2-a_{in}^2)\sigma(a) = 16 \pi a^2 \sigma_0 \left(\frac{a}{5 AU}\right)^{-\alpha} \left(\frac{M_p}{3M_{\odot}}\right)^{1/3} ,
\end{equation}
where $a_{out}$ and $a_{in}$ are the outer and inner radii of the
feeding zone, respectively. The planetesimals are assumed to be uniformly spread on either
side of the orbit to a distance $a_f$, which depends on the
eccentricity and inclination of the planetesimal orbits, as well as
the Hill sphere radius of the protoplanet $R_H$. We take $a_f\sim4R_H$, with $R_H=a\left(\frac{M_p}{3M_{\odot}}\right)^{1/3}$ and $M_p$ is the planet's mass. The inner and outer boundaries of the feeding zone are then taken to be  $a_{in}=a-a_f$ and $a_{out}=a+a_f$, respectively. We assume that planetesimals do not get in or out of the planetary feeding zone. A detailed discussion of how the feeding zone is defined can be found in Pollack et al. (1996). \par

The size of the feeding zone is proportional to $a^2$ while the solid surface density is proportional to $a^{-\alpha}$. As a result, the available mass of solids $M_{av}$ in the disk is proportional to $a^{2-\alpha}$ (equation (4)). Since we use $\alpha$ values smaller than 2, the available mass increases with radial distance. Figure 3 shows the available mass for accretion as a function of distance from the star for the three considered disk masses and density profiles. 
Although the available mass of heavy elements for accretion by a Jupiter-mass clump is found to increase with radial distance, it does not imply that this available mass can actually be accreted. The accretion rate is proportional to $\Omega$, the protoplanet's orbital frequency, which goes as $a^{-3/2}$, and to $\sigma$ the solid surface density ($\propto a^{-\alpha}$); both decrease with radial distance. As a result, the accretion timescale is significantly longer at the disk's outer regions. \par

The protoplanet can be enriched with heavy elements as long as the accretion process is efficient. While the protoplanet is extended (a few tenths of AU) it fills most of the area of its feeding zone, planetesimals are slowed down due to gas drag, and are absorbed by the protoplanet. In fact, both accretion and ejection occur all the time, but as long as the body is extended ejection is negligible. The accretion efficiency decreases with decreasing size of the protoplanet, and once the body is sufficiency small planetesimals no longer pass through its envelope, and instead of being accreted they get ejected from the protoplanet's vicinity. The radius of the protoplanet changes significantly once molecular hydrogen begins to dissociate and a dynamical collapse occurs. For a Jupiter-mass clump the dynamical collapse takes place after a few $10^5$ years (DeCampli and Cameron, 1978; Bodenheimer et al., 1980; paper I). After the collapse the body has much larger internal densities and temperatures, with a dimension of a couple of Jupiter radii. 
In addition, the protoplanet is expected to open a gap in the gas disk and undergo type II migration (Lin \& Papaloizou, 1985; Nelson et al., 2000). Once a gap is opened and inward migration occurs, planetesimals are mostly ejected by the migrating planet and accretion no longer dominates. In addition, it is unclear how planetesimals are distributed within the gap and how solids from the disk (or in the gap's vicinity if exist) interact with the protoplanet (Goldreich et al., 2004). The timescale for type II migration depends on the disk's viscosity and on the protoplanet's radial distance (D'Angelo et al., 2003). The migration timescale $\tau_{mig}$ decreases with increasing disk mass (massive disks are expected to be more viscous) and with decreasing radial distance ($\tau_{mig} \propto a^{1/2}$). For simplicity, for all cases considered here we take the maximum time for accretion as $10^5$ years. The accretion timescales for low mass disks and/or large radial distances are not expected to be more than a few $10^5$ years. For these cases, the accreted mass calculated here can be taken as a lower bound. 

\section{Results}

Figure 4a presents the captured mass in the first 10$^5$ years of the planetary evolution for all the cases considered. The black, blue and red curves are for disk masses of 0.1, 0.05, and 0.01 M$_{\odot}$, respectively. The solid, dashed, and dotted curves refer to density distributions proportional to $a^{-1/2}$, $a^{-1}$ and $a^{-3/2}$, respectively. 
As expected, higher solid surface density (more massive disks) results in larger enrichment in heavy elements. Lower $\alpha$ values allow heavy element capture at larger radial distances due to a more moderate decrease in density.  Although the available mass for accretion increases with radial distance, the location at which the maximum mass can be accreted does not exceed 15 AU. This is because the accretion time is significantly longer in the outer regions and the protoplanet cannot capture the entire available mass within $10^5$ years. 
The location of the maximum shifts slightly outwards with decreasing $\alpha$ since more mass is available at distant locations.  

\subsection{Jupiter}
We explore what disk configurations and formation locations lead to the heavy element enrichment as predicted by interior models of Jupiter. 
The heavy element mass in Jupiter's interior found by fitting its measured gravity field using various physical equations of state ranges from 20 to 40 M$_{\oplus}$ (Saumon and Guillot, 2004). 
Figure 4b focuses on the region of figure 4a in which the accreted heavy element mass is between 20 and 40 M$_{\oplus}$, as expected for Jupiter. The curves in this range represent the possible disk properties and formation radial distances that lead to a Jupiter-like heavy element abundance. \par

A disk mass of 0.01 M$_{\odot}$ is found to be too small to allow for a heavy element enrichment comparable with Jupiter's theoretical enrichment for all the considered density distributions. The accreted mass is found to be smaller than 10 M$_{\oplus}$ in all these cases. This is not surprising since this disk mass is smaller than that of the minimum-mass solar nebula (Weidenschilling, 1977).
For a disk mass of 0.05 M$_{\odot}$ with a radial density distribution proportional to $a^{-1/2}$ we find that the protoplanet can be enriched with 20 to 40 M$_{\oplus}$ if it is formed at radial distances between 7 and 10 AU, or 18 and 25 AU. 
For density distributions going as $a^{-1}$ and $a^{-3/2}$, we find that the protoplanet can accrete 20 - 40 M$_{\oplus}$ of heavy elements at locations between 5 - 9 or 17 - 22 AU, and 5 - 10 or 15 - 18 AU, respectively. 
A disk with a mass of 0.1 M$_{\odot}$ is massive enough to allow significant heavy element enrichment even at very large radial distances. 
However, due to the high solid surface density at radial distances smaller than $\sim$ 10 AU the enrichment in solids exceeds the amounts estimated for Jupiter, unless $\alpha$=1/2. Such high densities, however, may explain some of the observed 'over-enriched' extra-solar planets, as discussed in the following section.  An enrichment similar to that of Jupiter for a disk mass of 0.1 M$_{\odot}$ is achieved at radial distances of 5 - 6 or 25 - 30 AU, 22 - 29 AU, and 19 - 24 AU for $\alpha$ values of 1/2, 1 and 3/2, respectively.  Table 2 summarizes the disk configurations that lead to the enrichment expected for Jupiter. \par

Our results suggest that Jupiter could have formed at larger radial distances than its current location and still be enriched with the heavy element mass of interior models. Formation at larger radial distances is relevant for understanding the measured enrichment of 'metals' over solar composition in Jupiter's atmosphere (Owen et al., 1999; Mahaffy et al., 2000).  
The origin of heavy elements and volatiles in Jupiter's atmosphere is unknown and several authors have suggested different mechanisms to explain the similar enrichments of Ar, Kr, Xe, C, N, and S over solar composition in Jupiter's atmosphere (Owen et al., 1999; Gautier et al., 2001; Guillot and Hueso, 2006). 
Owen et al. (1999) suggested that Jupiter's atmosphere was enriched by impact of planetesimals formed beyond Jupiter's location where the temperatures were significantly lower ($\sim$ 25 K) and trapping of volatile gases in amorphous ice was possible. This scenario can explain why argon, krypton and xenon are enriched in similar proportions as the other heavy elements (Owen et al., 1999). Since the formation of the accreted bodies must occur at significantly lower temperatures than expected at 5 AU, Owen et al. (1999) suggested that Jupiter was formed at a low temperature region and subsequently  migrated to its present position. If Jupiter indeed formed farther out, the accreted mass would naturally consist of amorphous ice in which noble gases can be trapped. Jupiter's formation at a large radial distance by a gravitational instability not only leads to the required amount of heavy elements in its interior but it may also explain the enrichment of volatiles in its atmosphere. 

\subsection{Extra-solar Planets}

Observations of a transiting extra-solar planet provide estimates of the planet's mass and radius and therefore, its mean density.
Knowledge of the mean density constrains the mass of heavy elements within the planet's interior. Detailed studies of transiting giant extra-solar planets suggest that the mass of heavy elements ranges from 0 up to 100 M$_{\oplus}$ (Guillot et al., 2006; Guillot, 2007). 
Our calculations show that the accreted heavy element mass changes significantly, depending on the disk's mass and its density profile. 
Even-though the protoplanet starts with a solar composition it can end up with a non-solar abundance.  
The parameter which determines the possible enrichment is the solid surface density at the location in which the planet forms. For the disk parameters considered here, the heavy element enrichment of a Jupiter-mass protoplanet is found to range from about 1 to 110 M$_{\oplus}$, resulting in final planetary masses between 1 and 1.35 masses of Jupiter. \par

In this paper the stellar metallicity was taken as a constant; therefore the solid surface density increases with increasing disk mass. However, surface densities change with stellar metallicity. Disks around high metallicity stars 
can have relatively low masses and still have high solid surface densities leading to a substantial enrichment of 'metals' (Doson-Robinson and Bodenheimer, 2009). Since we find that higher surface density (metallicity) leads to larger enrichment in heavy elements, our findings are consistent with the correlation between the mass of heavy elements in a giant planet and stellar metallicity (Guillot et al., 2006; Guillot, 2007). In addition, our work offers a mechanism that can lead to both highly enriched planets, such as the extra-solar planet HD 149026b which is predicted to contain about 60 to 80 M$_{\oplus}$ of heavy elements (Sato et al., 2005; Baraffe et al., 2008), and to giant planets with very low densities, such as TrES-4b (Mandushev et al. 2007; Guillot, 2007). \par 

\section{Summary and Conclusions}

We present a calculation of heavy element enrichment via planetesimal capture by a Jupiter-mass clump created by disk instability. 
The accreted mass is computed considering different formation radial distances for the protoplanet (5 - 30 AU), and various disk environments. We consider three disk masses (0.1, 0.05, and 0.01 M$_{\odot}$) and solid surface density distributions ($\propto a^{-1/2},a^{-1}$, and $a^{-3/2}$).  
Massive disks have high surface densities allowing for a significant enrichment of heavy elements. Increasing values of $\alpha$ (increasing the density function's steepness) lead to lower heavy element enrichment in the outer part of the disk.\par

The available mass for accretion increases with radial distance. The accretion rate, however, decreases with radial distance due to its dependence on both the surface density and orbital frequency, both of which decrease with radial distance.  
The maximum time for accretion is taken to be $10^5$ years, the period in which planetesimal accretion is efficient. The exact amount of heavy elements accreted will depend on both the disk properties (mass, density distribution, viscosity, etc.) and the planetary contraction rate. The contraction rate can be slower when the effects of planetesimal accretion and stellar radiation are included in the evolution calculation. 
At small radial distances these effects are unimportant because the accretion timescale is shorter or comparable to the $10^5$ years considered. At large radial distances the stellar radiation effect is small but since the accretion time is significantly longer a slower contraction provides more time for accretion, and therefore increases the captured mass. 
As a result, the accreted heavy element mass for large radial distance presented here is a lower bound. \par
 
For the disk parameters considered here we find that heavy element enrichment in a Jupiter-mass protoplanet can range from about 1 to 110 M$_{\oplus}$. The result shows that protoplanets with similar initial masses (one Jupiter-mass in our case) can end up having significantly different compositions, as well as final masses. Since both the planetary contraction rate and accretion rate change with planetary mass, protoplanets with varying initial masses will have different enrichments. 
The planetary enrichment also depends on the solid surface density which changes with the disk's mass, size (outer boundary), and its density profile. The enrichments presented here, are therefore consequences of the values chosen to define the disk properties. Once disk configurations can be described more accurately, the planet's enrichment could be better constrained. Finally, a planet's enrichment also changes with the planetesimal properties (size, composition, velocities), as shown in paper I. The capture efficiency (cross-section) decreases with  increasing planetesimal size, density, and random velocity. Our calculations can be repeated using different disk environments, planetary masses,  and planetesimal properties. \par 

We conclude that Jupiter could have formed at larger radial distances than its current location (the exact radial distance depends on the assumed disk parameters) and still accrete between 20 to 40 M$_{\oplus}$ of heavy elements, the heavy element mass predicted by interior models (Saumon and Guillot, 2004). Since the disk's temperature decreases with increasing radial distance, at larger distances planetesimals can consist of amorphous ice in which volatiles can be trapped. Formation at more distant regions explains the enrichment of heavy elements and volatiles measured in Jupiter's atmosphere (Owen et al., 1999). However, since the disk's temperature can increase with increasing surface density, a more detailed analysis is needed to evaluate the temperature profile, and the radial distances required to allow the existence of amorphous ice and efficient trapping of volatiles. In such an investigation larger radial distances than considered here may be required. \par  

Our work suggests that the final composition of the protoplanet can change considerably depending on the planetary 'birth environment'. The large variation we derive in heavy element enrichment explains the different compositions of observed extra-solar gas giant planets (Guillot, 2007). Since higher surface density leads to larger heavy element enrichments, our model results are found to be consistent with the correlation between heavy element enrichment and stellar metallicity (Guillot et al., 2006). \par 

We conclude that the disk instability model can lead to both giant planets with nearly solar compositions and planets which are significantly enriched with heavy elements. Our findings support the disk instability model of gas giant planet formation. 
\par


\acknowledgments
R. H thanks Alan Boss, Nadar Haghighipour, Brad Hansen, and Morris Podolak for fruitful discussions and helpful suggestions. 
R. H. acknowledges support from NASA through the Southwest Research Institute. G. S. acknowledges support from the NASA PGG and PA programs. 

\section*{References} 
\noindent Andrews, S. M., Hughes, A. M., Wilner, D. J., Qi, C., The Structure of the DoAr 25 Circumstellar Disk. ApJ, 678, (2008), pp. L133--L136.\\
Andrews, S. M.; Williams \& J. P., High-Resolution Submillimeter Constraints on Circumstellar Disk Structure. ApJ, 659, (2007), pp. 705--728.\\ 
Baraffe, I., Chabrier, G., \& Barman, T., Structure and Evolution of Super-Earth to SuperJupiter Exoplanets. I. Heavy Element Enrichment in the Interior. A\&A, 482, (2008), pp. 315--332.\\ 
Bodenheimer, P., Grossman, A. S., Decampli, W. M., Marcy, G. \& Pollack, J. B., Calculations of the evolution of the giant planets, Icarus, 41, (1980), pp. 293--308.\\
 Boss, A. P., Giant planet formation by gravitational instability, Science, 276, (1997), pp. 1836--1839.\\
Boss, A. P. Analytical Solutions for Radiative Transfer: Implications for Giant Planet Formation by Disk Instability, ApJ, (2009), in press.\\
D'Angelo, G, Henning, T. \& Kley, W.	Thermohydrodynamics of Circumstellar Disks with High-Mass Planets, ApJ, 599 (2003), pp. 548--576.\\
Decampli, W. M. \& Cameron, A. G. W., Structure and evolution of isolated giant gaseous protoplanets, Icarus, 38 (1979), pp. 367--391.\\
Dodson-Robinson, S. E. \& Bodenheimer, P., Discovering the Growth Histories of Exoplanets: The Saturn Analog HD 149026b. ApJ Letters, (2009) submitted. (arXiv:0901.0582)\\
Dodson-Robinson, S. E., Bodenheimer, P., Laughlin, G., Willacy, K., Turner, N. J. \& Beichman, C. A., Saturn Forms by Core Accretion in 3.4 Myr. ApJ, 688, (2008), pp. L99--L102. \\
Dodson-Robinson, S. E., Willacy, K., Bodenheimer, P., Laughlin, G., Turner, N. J., \& Beichman, C., Ice Lines, Planetesimal Composition and Solid Surface Density in the Solar Nebula. Icarus, in press, (2008), (arXiv:0806.3788).\\
Durisen, R. H., Boss, A. P., Mayer, L., Nelson, A. F., Quinn, T., \& Rice, W. K. M., Gravitational Instabilities in Gaseous Protoplanetary Disks and Implications for Giant Planet Formation. in Protostars and Planets V, ed. B. Reipurth, D. Jewitt, \& K. Keil, (2007), pp. 607--622.\\ 
Eisner, J. A., Plambeck, R. L., Carpenter, John M., Corder, S. A. \& Qi, C.; Wilner, D., Proplyds and Massive Disks in the Orion Nebula Cluster Imaged with CARMA and SMA. ApJ, 683, (2008), pp. 304--320. \\
Eisner, J. A. \& Carpenter, J. M., Massive Protoplanetary Disks in the Trapezium Region. ApJ, 641, (2006), pp. 1162--1171.\\
Eisner, J. A., Hillenbrand, L. A., White, R. J., Akeson, R. L. \& Sargent, A. I., Observations of T Tauri Disks at Sub-AU Radii: Implications for Magnetospheric Accretion and Planet Formation. ApJ, 623, (2005), pp. 952--966.\\ 
Gautier, D.,  Hersant, F., Mousis, O. \&Lunine, J. I.,  Enrichments in Volatiles in Jupiter: A New Interpretation of the Galileo Measurements. ApJ, (2001), 550, pp. L227--L230.\\
Greenzweig, Y., Lissauer, J.J., Accretion rates of protoplanets. Icarus, 87, (1990), pp. 40--77. \\
Goldreich, P., Lithwick, Y. \& Sari, R., Final Stages of Planet Formation. ApJ, 614, (2004), pp. 497--507.\\
Guillot, T., Santos, N. C., Pont, F., Iro, N.,  Melo, C. \& Ribas, I. A correlation between the heavy element content of transiting extrasolar planets and the metallicity of their parent stars, A\&A, 453, (2006), pp. L21--L24.\\
Guillot, T. \& Hueso, R., The composition of Jupiter: sign of a (relatively) late formation in a chemically evolved protosolar disc. MNRAS, 367, (2006), L47--L51. \\
Guillot, T., The composition of transiting giant extra-solar planets. Physica Scripta, 130, (2007), pp. 014023--014029.\\
Helled, R., Podolak, M. \& Kovetz, A., Planetesimal capture in the disk instability model. Icarus 185 (2006), pp. 64--71.\\
Helled, R., Podolak, M. \& Kovetz, A., Grain sedimentation in a giant gaseous protoplanet. Icarus, 195 (2008), pp. 863--870. \\
Hubickyj, O., Bodenheimer, P., \& Lissauer, J. J. Accretion of the gaseous envelope of Jupiter around a 5Ð10 Earth-mass core, Icarus 179 (2005), pp. 415--431.\\
Lissauer, J. J. \& Stevenson, D. J., Formation of Giant Planets. PPV, B. Reipurth, D. Jewitt, \& K. Keil (eds.), University of Arizona Press, Tucson (2007), 951, pp. 591--606.\\
Lissauer, J. J., Hubickyj, O., D'Angelo, G. \& Bodenheimer, P., Models of Jupiter's growth incorporating thermal and hydrodynamic constraints. Icarus, 199, (2009), pp. 338--350.\\
Mahaffy, P. R., Niemann, H. B., Alpert, A., Atreya, S. K., Demick, J., Donahue, T. M., Harpold, D. N. \& Owen, T. C., 
Noble gas abundance and isotope ratios in the atmosphere of Jupiter from the Galileo Probe Mass Spectrometer 
JGR, 105, (2000), pp. 15061--15072.\\
Mandushev, G., OÕDonovan, F. T., Charbonneau, D., Torres, G., Latham, D. W., Bakos, G. A., Dunham, E. W., Sozzetti, A., Fernandez, J. M., Esquerdo, G. A., Everett, M. E., Brown, T. M., Rabus, M., Belmonte, J. A. \& Hillenbrand, L. A., TrES-4: A Transiting Hot Jupiter of Very Low Density. ApJ, 667, (2007), pp. L195--L198.\\
Marois, C., Macintosh, B. , Barman, T.,  Zuckerman, B., Song, I., Patience, J., Lafreniere, D. \& Doyon, R., Direct Imaging of Multiple Planets Orbiting the Star HR 8799. Science, 322, (2008), pp. 1348--1352.\\\
Mayer, L., Quinn, T., Wadsley, J.,  \& Stadel, J. Formation of Giant Planets by Fragmentation of Protoplanetary Disks, Science, 298 (2002),  pp. 1756--1759.\\
Militzer, B., Hubbard, W. B., Vorberger, J., Tamblyn, I. \& Bonev, S. A. A Massive Core in Jupiter Predicted from First-Principles Simulations, ApJL, 688, (2008), pp. L45--L48.\\
Nelson, R. P., Papaloizou, J. C. B.,  Masset, F. \& Kley, W., The migration and growth of protoplanets in protostellar discs. MNRAS, 318, ( 2000), pp. 18--36.\\
Nettelmann, N., Holst, B., Kietzmann, A., French, M., Redmer, R. \& Blaschke, D., Ab Initio Equation of State Data for Hydrogen, Helium, and Water and the Internal Structure of Jupiter. ApJ, 683, (2008), pp. 1217--1228.\\ 
Owen, T., Mahaffy, P., Niemann, H. B. Atreya, S., Donahue, T., Bar-Nun, A. \& de Pater, I.,  A low-temperature origin for the planetesimals that formed Jupiter. Nature, 402, (1999), pp. 269--270.\\
Podolak, M., Pollack, J. B., \& Reynolds, R. T. The interaction of planetesimals with protoplanetary atmospheres, Icarus 73 (1987), pp. 163--179.\\
Pollack, J. B., Hubickyj, O., Bodenheimer, P., Lissauer, J. J., Podolak, M., \& Greenzweig, Y. Formation of the giant planets by concurrent accretion of solids and gas, Icarus 124 (1996), pp. 62--85.\\
Rafikov, R. R. Convective Cooling and Fragmentation of Gravitationally Unstable Disks, ApJ, 662 (2007),  pp. 642--650.\\ 
Ruden, S. P. \& Lin, D. N. C., The global evolution of the primordial solar nebula. ApJ, 308, (1986), pp. 883--901.\\ 
Sato B., Fischer D. A., Henry G. W., Laughlin G., Butler R. P., et al., The N2K Consortium. II. A Transiting Hot Saturn Around HD 149026 With a Large Dense Core. ApJ, 633, (2005), pp. 465--473.\\ 
Saumon, D. \& Guillot, T., Shock compression of deuterium and the interiors of Jupiter and Saturn, ApJ, 609 (2004), pp. 1170--1180.\\
Shu, F., Najita, J., Ostriker, E., Wilkin, F., Ruden, S., \& Lizano, S.,Magnetocentrifugally driven flows from young stars and disks. 1: A generalized model. ApJ, 429, (1994), pp. 781--796.\\
Weidenschilling, S. J., The distribution of mass in the planetary system and solar nebula. Astrophysics and Space Science, 51, (1977), pp. 153--158. \\
Young, R.E., The Galileo probe: How it has changed our understanding of Jupiter. New Astron. Rev., 47, (2003), pp. 1--51.

\clearpage

\begin{figure}[h!]
   \centering
    \includegraphics[width=3.4in]{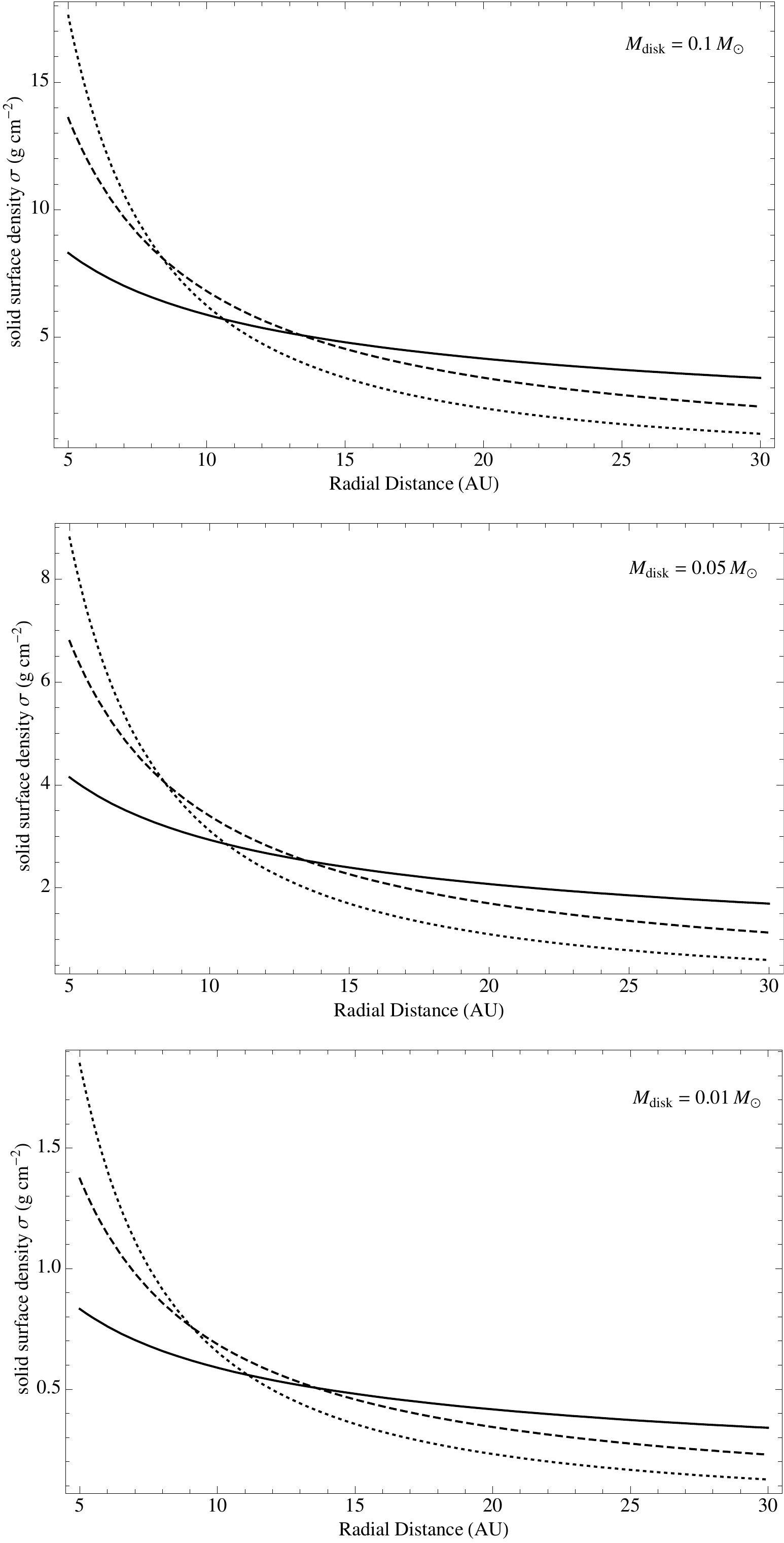}
    \caption[err]{Solid surface density as a function of radial distance for the three different disk masses (0.1, 0.05, and 0.01 M$_{\odot}$). The solid, dashed, and dotted curves represent surface density distributions with $\alpha$ = 1/2, 1 and 3/2, respectively. }
\end{figure}

\begin{figure}[h!]
   \centering
    \includegraphics[width=5.in]{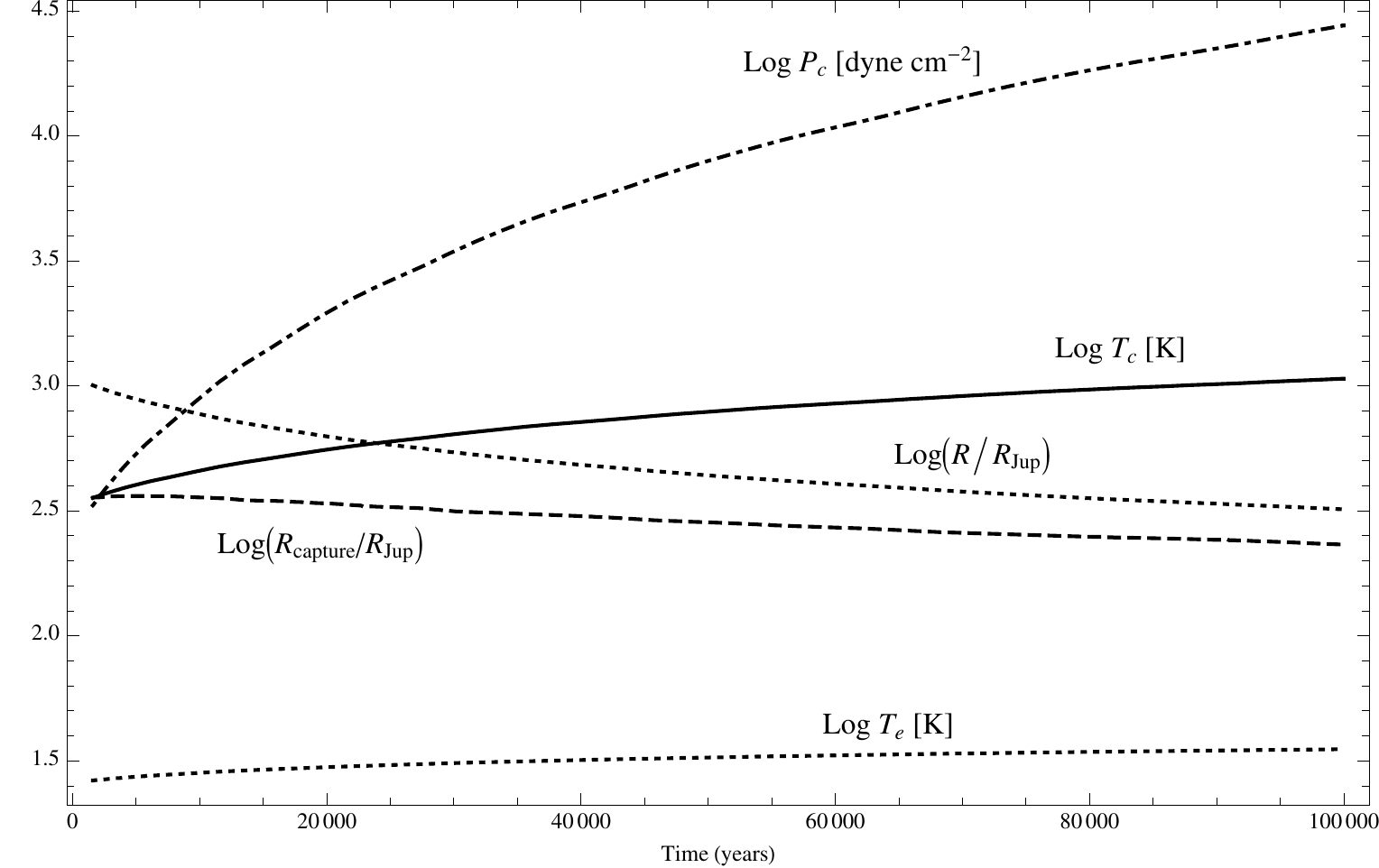}
    \caption[err]{Central pressure $P_c$, central temperature $T_c$, radius $R$, capture radius $R_{capture}$(for 1 km radius planetesimals made of ice and rock, and a random velocity of 1 km s$^{-1}$), and effective temperature $T_e$ of a Jupiter-mass protoplanet as a function of time.}
\end{figure}

\begin{figure}[h!]
   \centering
    \includegraphics[width=3.4in]{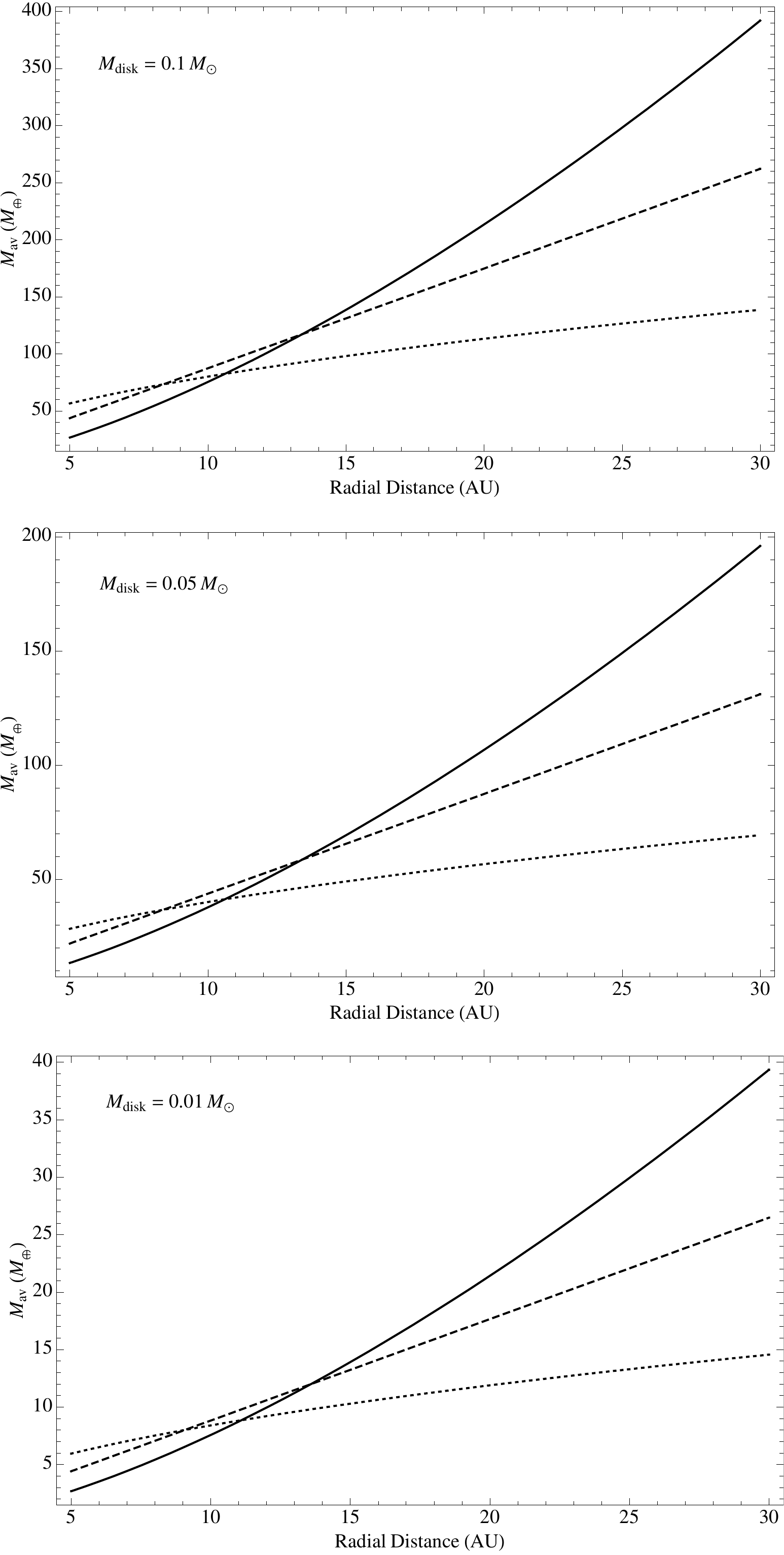}
    \caption[err]{Available mass for capture as a function of radial distance for three different disk masses (0.1, 0.05, and 0.01 M$_{\odot}$). The solid, dashed, and dotted curves represent surface density distributions with $\alpha$ = 1/2, 1 and 3/2, respectively.}
\end{figure}

\begin{figure}[h!]
   \centering
    \includegraphics[width=4.4in]{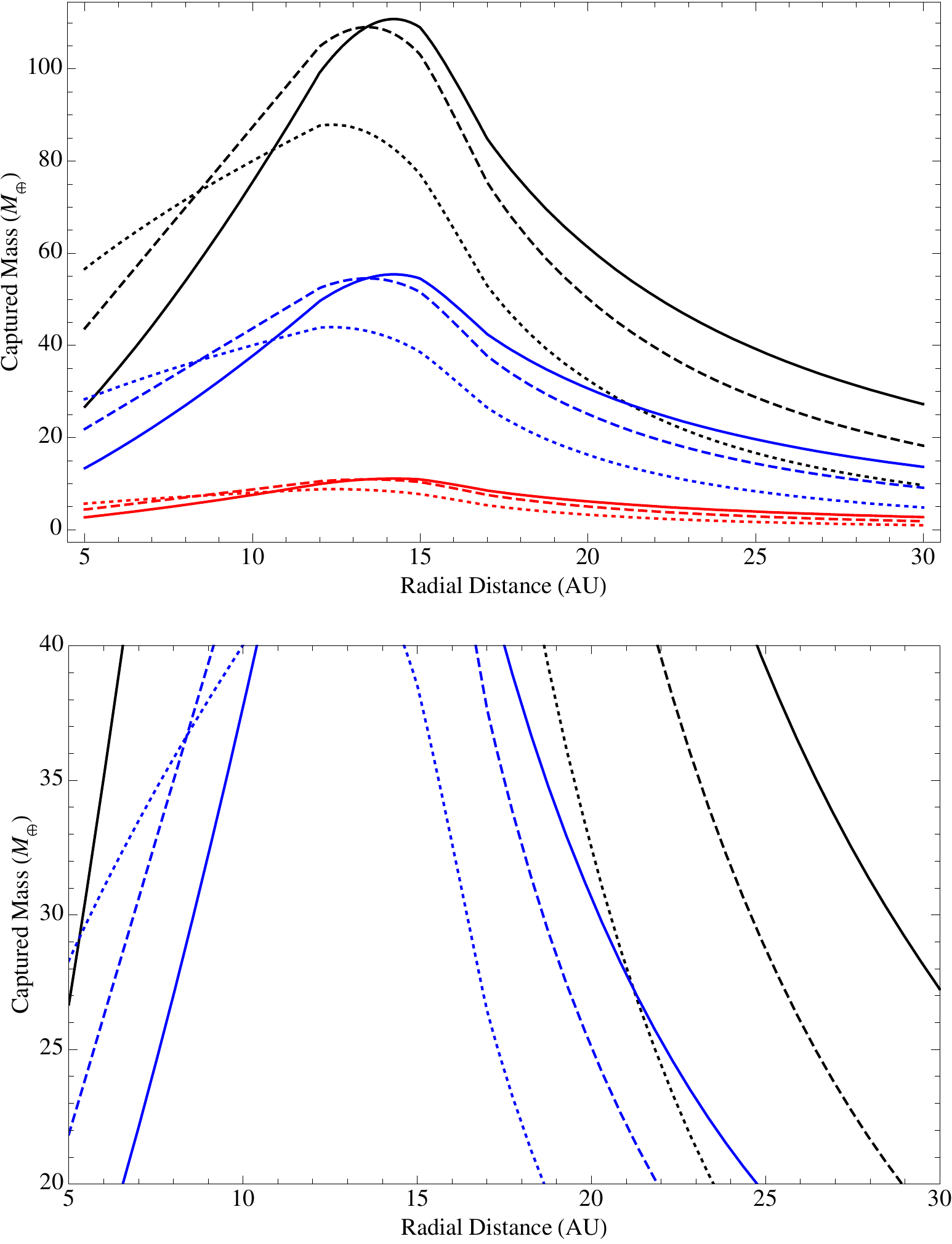}
    \caption[err]{(a) Captured mass as a function of radial distance for three different disk masses and density distributions. The black, blue, and red curves present the results for disk masses of 0.1, 0.05, and 0.01 M$_{\odot}$, respectively. The solid, dashed, and dotted curves are for density distributions with $\alpha$ = 1/2, 1 and 3/2, respectively. If accretion times exceed 10$^5$ years, the results provide lower bounds on the amount of heavy element enrichment. (b) Configurations which provide the heavy element enrichment between 20 and 40 M$_{\oplus}$ as predicted by Jupiter interior models.}
\end{figure}

\clearpage

\begin{table}[h!]
\centering
{\renewcommand{\arraystretch}{0.9}
\begin{tabular}{|c|c|c|c|}
\hline
Disk Mass & $\sigma_0(\alpha=1/2)$ & $\sigma_0(\alpha=1)$ &$\sigma_0(\alpha=3/2)$ \\
(M$_{\odot}$) & (g cm$^{-2}$) & (g cm$^{-2}$) &(g cm$^{-2}$) \\
\hline
0.1  & 8.29 & 13.59 &17.61 \\
0.05 & 4.15  & 6.8 &8.8 \\
0.01 & 0.83 & 1.36 &1.76 \\
\hline
\end{tabular}
}
\caption{{\small Solid surface densities at 5 AU ($\sigma_0$) for the disk masses and $\alpha$ values used in this work.}} \label{tab:1}
\end{table}

\vskip 4cm
 
\begin{table}[h!]
\centering
{\renewcommand{\arraystretch}{0.9}
\begin{tabular}{|c|c|c|}
\hline
Disk Mass (M$_{\odot}$) & $\alpha$ & Radial Distance (AU) \\
\hline
0.05 & 1/2& 7-10; 18-25 \\
0.05 & 1  & 5-9; 17-22\\
0.05 & 3/2 & 5-10; 15-18\\
0.1 & 1/2 & 5-6; 25-30 \\
0.1 & 1 & 22-29\\
0.1 & 3/2 & 19-24 \\
\hline
\end{tabular}
}
\caption{{\small Disk configurations that lead to heavy element enrichment between 20 and 40 M$_{\oplus}$. The third column gives the radial distances in which Jupiter could have formed for a specific disk model. }} \label{tab:2}
\end{table}

\end{document}